\documentstyle[psfig,subfigure,mncite]{mn}
\begin{document}
    
\title[Modelling surface magnetic field evolution on AB Dorad\^{u}s]
{Modelling surface magnetic field evolution on AB Dorad\^{u}s due to
diffusion and surface differential rotation}

\author[G.R.~Pointer et al.]{
G.R.~Pointer$^1$,
M.~Jardine$^1$
\thanks{E-mail: moira.jardine@st-and.ac.uk},
A.~Collier Cameron$^1$ and 
J-F.~Donati$^2$\\
$^1$School of Physics and Astronomy, University of St Andrews, St
Andrews, Fife, KY16 9SS, Scotland\\
$^2$Laboratoire d'Astrophysique, Observatoire Midi-Pyr\'{e}n\'{e}es,
F-31400
Toulouse, France}

\date{}
\maketitle

\begin{abstract}
From Zeeman Doppler images of the young, rapidly-rotating K0 dwarf AB
Dorad\^{u}s, we have created a potential approximation to the observed
radial magnetic field and have evolved it over 30 days due to the
observed surface differential rotation, meridional flow and various
diffusion rates.  Assuming that the dark polar cap seen in Doppler
images of this star is caused by the presence of a unipolar field, we
have shown that the observed differential rotation will shear this
field to produce the observed high-latitude band of unidirectional
azimuthal field.  By cross-correlating the evolved fields each day
with the initial field we have followed the decay with time of the
cross-correlation function.  Over 30 days it decays by only 10\%. 
This contrasts with the results of \scite{barnes98alpper} who show
that on this timescale the spot distribution of He699 is uncorrelated. 
We propose that this is due to the effects of flux emergence changing
the spot distributions.
\end{abstract} 

\begin{keywords}
 stars: activity --  
 stars: imaging --
 stars: individual: AB Dor --
 stars: magnetic fields --
 stars: rotation --
 stars: spots 
\end{keywords}

\section{Introduction}

AB Dorad\^{u}s is a K0 dwarf, originally discovered as a bright X-ray
source by \scite{pakull81} and subsequently observed by ROSAT
\cite{kurster97} and {\em Beppo}SAX \cite{maggio00}.  Its photometric
variability is believed to be due to starspots (\pcite{anders90},
\pcite{innis88}) and this, combined with its brightness ($V\simeq
6.8-7.0$) and rapid rotation (P=0\fd514) have made it an attractive
candidate for Doppler imaging (\pcite{kuerster94},
\pcite{cameron94doppler}, \pcite{cameron95doppler},
\pcite{unruh95doppler},\pcite{donati97doppler},
\pcite{donati99doppler},\pcite{cameron99musicos}).  From observations
of the lithium line at 6708\AA, \scite{rucinski82} suggested it was a
post-T Tauri star.  According to HIPPARCOS data it is 14.94$\pm$0.12
pc away. \scite{cameron97rst137b} inferred an age of $\sim$2-3x10$^{7}$ 
year and its common-proper-motion companion, the M dwarf Rst 137b.

AB Dor is of interest for a variety of reasons.  The most important,
for the purposes of this paper, is that Zeeman Doppler images have
been obtained in 3 consecutive years: 1995 Dec
7-13\cite{donati97doppler}; 1996 Dec 23-29\cite{donati99doppler}; and
1998 Jan 10-15.  These studies reveal that the radial field has at
least 12 regions of opposite polarities at intermediate to high
latitude, which are approximately regularly spaced in longitude
together with a unidirectional ring of azimuthal field at 70-80\degr
indicating an underlying large-scale toroidal field
\cite{donati99doppler}.

AB Dor exhibits flaring X-ray emission (\pcite{cameron88xray},
\pcite{schmitt98}) and there is indirect evidence from radio
observations at 3,6,13 and 20 cm that the radio emission is highly
directive and suggests synchrotron radiation \cite{lim94}.  The star
is surrounded by a system of circumstellar prominences which can be
observed as absorption transients in optically thick low-excitation
lines e.g.\ H Balmer, Ca{\sc ii} and Mg{\sc ii} when the prominences
cross the line of sight
(\pcite{cameron89cloud},\pcite{cameron89eject},\pcite{cameron90masses}).
These prominences are trapped by the stellar magnetic field at, or
beyond, the point of centrifugal balance.  Their presence demonstrates
that the corona is highly structured even as far out as 3-5$R_{*}$
\cite{jardine96flow}.

Despite the rapid rotation, the differential rotation has been
measured by cross-correlation of Zeeman Doppler images secured a few
days apart to be close to solar, with the equator lapping the poles in
$\sim$110$^{d}$ ({\em cf.} 120$^{d}$ in the solar case)
\cite{donati97doppler,donati99doppler}.  On longer timescales,
however, \scite{barnes98alpper} have shown that the spot distribution
of the similar young rapid rotator He 699 becomes uncorrelated after 30
days.

The purpose this paper is to investigate the effect of diffusion and
differential rotation on the evolution of AB Dorad\^{u}s's magnetic
field.  We aim to find out whether shearing at the edge of a unipolar
cap can produce the observed ring of unidirectional azimuthal field. 
We also seek to determine the lifetimes of surface magnetic features
subject to diffusion and differential rotation.
 
\section{Temporal evolution of the magnetic field}

We are using a code originally developed by \scite{vanballegooijen98}
to study the formation of filament channels on the Sun.  It can also
be used to study the field of AB Dorad\^{u}s because we have
high-resolution magnetic maps (2\fdg9 at the equator) and the
differential rotation is similar to the solar value.  The code takes
the observed surface radial component of the field and calculates a
potential field from this, and then evolves the calculated magnetic
field due to the effects of differential rotation and diffusion.

\scite{jardine99potent} have demonstrated that for latitudes below
about 60\degr the field is well-represented by a potential
approximation.  We anticipate that departures from a potential field
caused by the shearing effect of the differential rotation will appear
at high latitudes near the edge of the unipolar cap.  By fitting the
Stokes V profiles with both potential \cite{hussain2001technique} and
non-potential \cite{hussain2001cs12} field models, it is possible to
show that any currents are concentrated close to the pole.

\subsection{The scalar magnetic potential, $\psi$}

If we assume that the field is potential, then we can write {\bf B} in terms of a flux
function $\psi$, with
$${\bf B}=-{\bf \nabla}\psi,$$ which in spherical co-ordinates gives
$$B_{r}=-\frac{\partial\psi}{\partial r},$$

$$B_{\theta}=-\frac{1}{r}\frac{\partial\psi}{\partial\theta}
$$
and
$$
B_{\phi}=-\frac{1}{r\sin\theta}\frac{\partial\psi}{\partial\phi}.
$$
Here $\psi$ satisfies Laplace's equation $\nabla^{2}\psi=0$ which can be
expressed as
\begin{equation}
\frac{1}{r^{2}}\frac{\partial}{\partial
r}(r^{2}\frac{\partial\psi}{\partial
r})+\frac{1}{r^{2}\sin\theta}\frac{\partial}{\partial\theta}(\sin\theta
\frac{\partial\psi}{\partial\theta})+\frac{1}{r^{2}\sin^{2}\theta}
\frac{\partial^{2}\psi}{\partial\phi^{2}}=0.
\end{equation}
A separable solution for $\psi$ can be found
$$
\psi(r,\theta,\phi)=\sum_{l=1}^{N}\sum_{m=-l}^{l}\psi_{lm}(r)P_{lm}(\theta)
e^{im\phi},
$$
where $P_{lm}$ are the associated Legendre functions and
$$
\psi_{lm}(r)=a_{lm}r^{l}+b_{lm}r^{-(l+1)}.
$$
We chose to truncate the series at $N=63$, corresponding to the
maximum resolution of the reconstructed field images.  We clearly need
two boundary conditions to determine $\psi$.  We chose to specify as
one boundary condition that at some distance from the star (the {\em
source surface}, $r_{s}\approx 5.1R_{*}$), the field is radial and so
$B_{\theta}(r_{s})=B_{\phi}(r_{s})=0$ \cite{schatten69}.  This mimics
the stellar wind.

Since most stellar prominences form at around the corotation radius
(2.7$R_{*}$), we know that a significant fraction of the field is
closed at that radius.  Hence we choose $r_{s}=5.1R_{*}$.
We then have
$$
a_{lm}r_{s}^{l-1}+b_{lm}r_{s}^{-l-2}=0,
$$
equivalent to 
$$
\psi_{lm}=0.
$$
As a second boundary condition we impose the radial field at the
surface to be the observed radial field.  We can then express the
magnetic field in terms of the two-dimensional Fourier coefficients
$B_{lm}$, where 
$$
B_{lm}(R_{*})=2\pi \int_{0}^{\pi}
B_{m}(\theta)P_{lm}(\theta)\sin
\theta d\theta
$$
so 
$$
B_{lm}(R_{*})=-la_{lm}R_{*}^{l-1}+(l+1)b_{lm}R_{*}^{-l-2}.
$$
The function $B_{m}(\theta)$ is derived from a fast Fourier transform
performed latitude-by-latitude on the observed radial field
$B_{r}(R_{*},\theta,\phi)$.

Once the field is evolved due to diffusion and differential rotation,
it is not necessarily potential, although it can still be expressed as
a sum of spherical harmonics.  The field components are then expressed
in terms of the functions
$$
J=\sum_{l=1}^{N}\sum_{m=-l}^{l}J_{lm}P_{lm}(\theta)e^{im\phi}
$$ 
and
$$
A=\sum_{l=1}^{N}\sum_{m=-l}^{l}A_{lm}P_{lm}(\theta)e^{im\phi}
$$ 
where
$$
J=\frac{1}{\sin\theta}\left[\frac{\partial}{\partial\theta}(\sin\theta
B_{\phi})-\frac{\partial B_{\theta}}{\partial\phi}\right]
$$ 
is the radial component of the current and
$$
A=\frac{1}{\sin\theta}\left[\frac{\partial}{\partial\theta}(\sin\theta
B_{\theta})+\frac{\partial B_{\phi}}{\partial\phi}\right]
$$ 
is the 2-dimensional divergence.

\subsection{Evolving the field using the induction equation}

\begin{figure*}
\begin{tabular}{cc}
\psfig{file=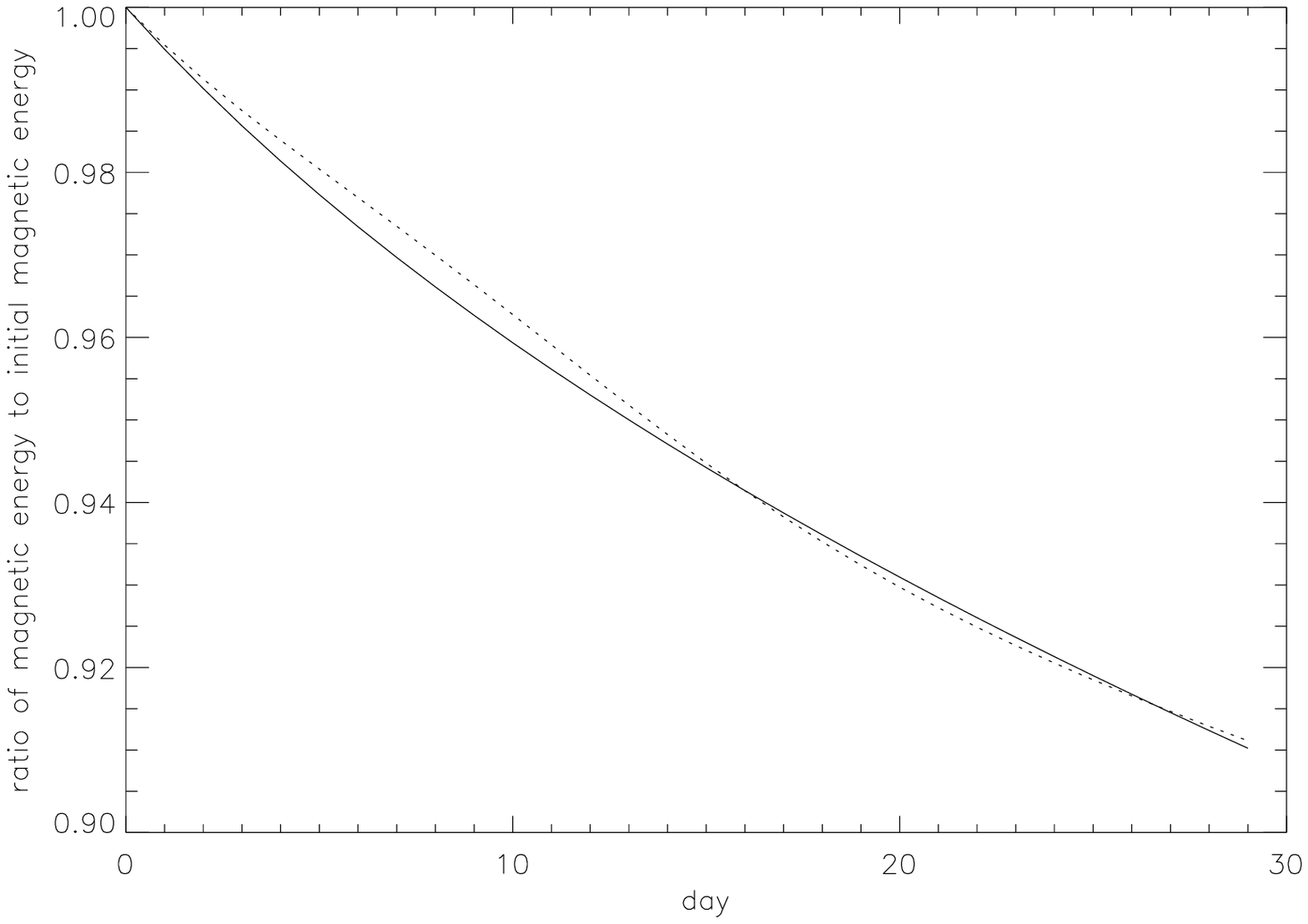,angle=0,width=7.0cm} &
\psfig{file=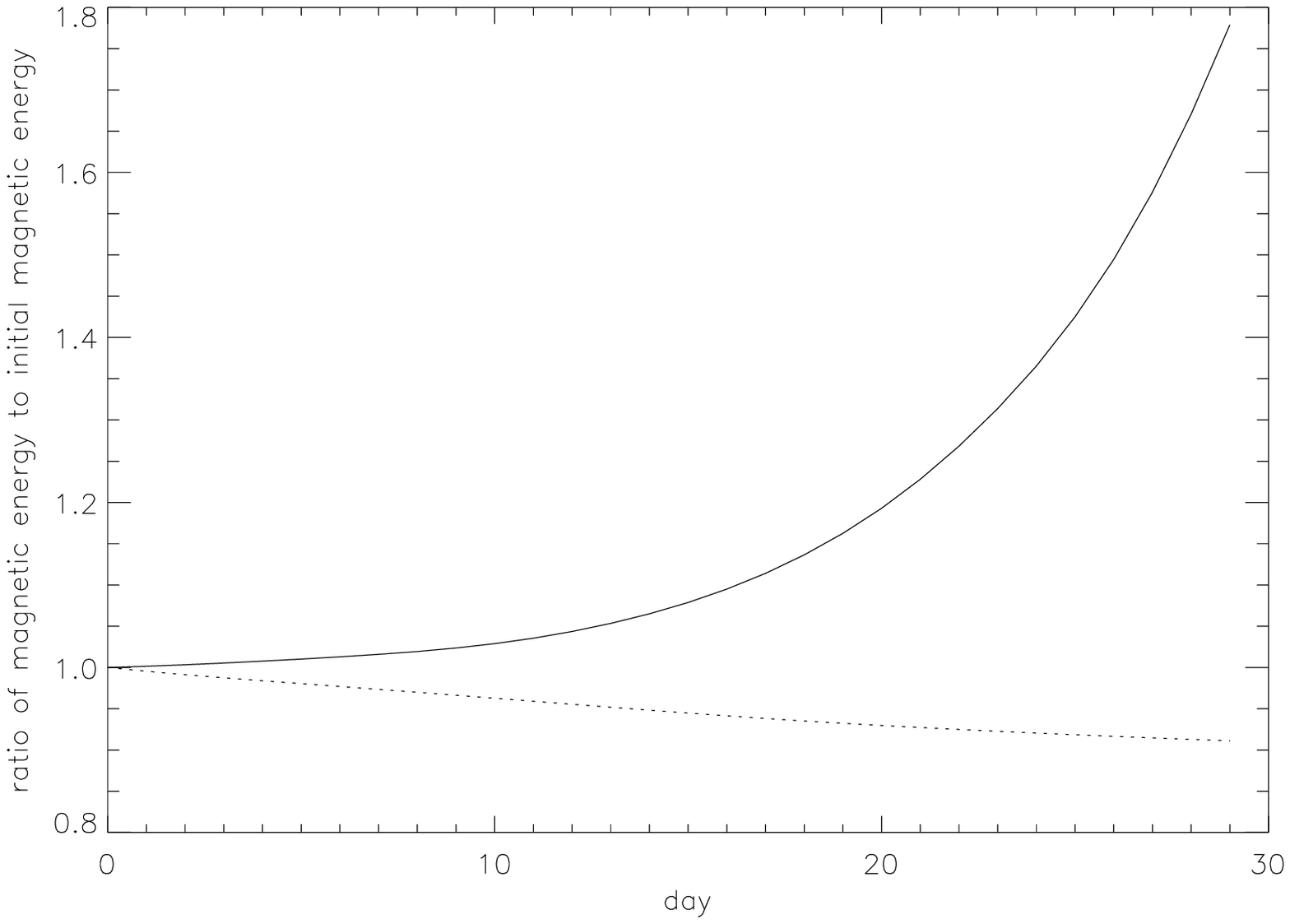,angle=0,width=7.0cm} \\
\end{tabular}
\caption{This shows how the magnetic energy varies as the field evolves
for 1998. The left hand diagram is for the case when
$\eta$=250km$^{2}$s$^{-1}$, with differential rotation (solid line) and
without (dotted line), and the right hand diagram
compares the case where
there is no diffusion, i.e.\ $\eta=0$ (solid line) and where
$\eta$=250km$^{2}$s$^{-1}$
(dotted line).} 
\label{magen} 
\end{figure*}

\begin{figure*}
\begin{tabular}{cc}
\psfig{file=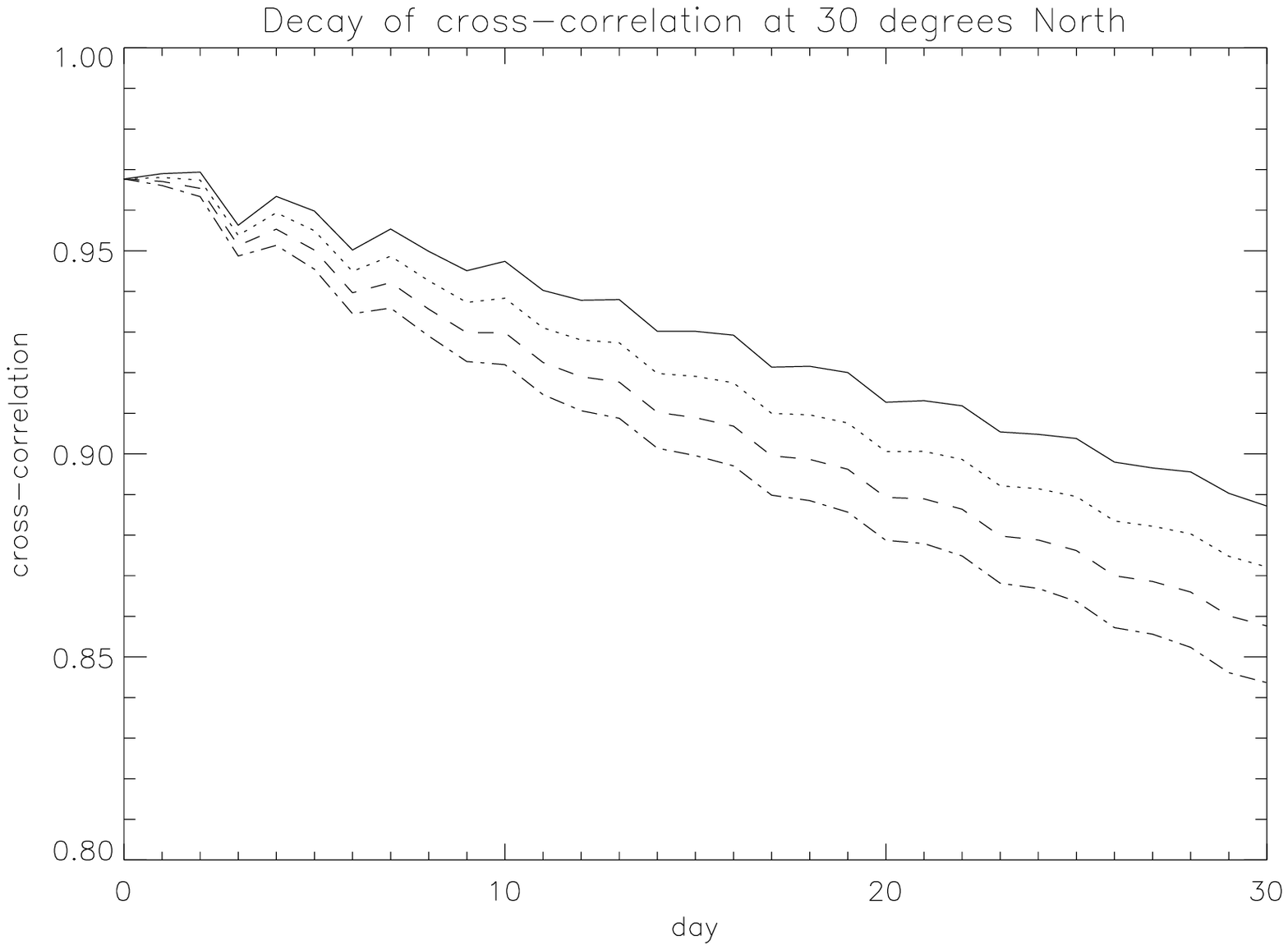,angle=0,width=7.0cm} &
\psfig{file=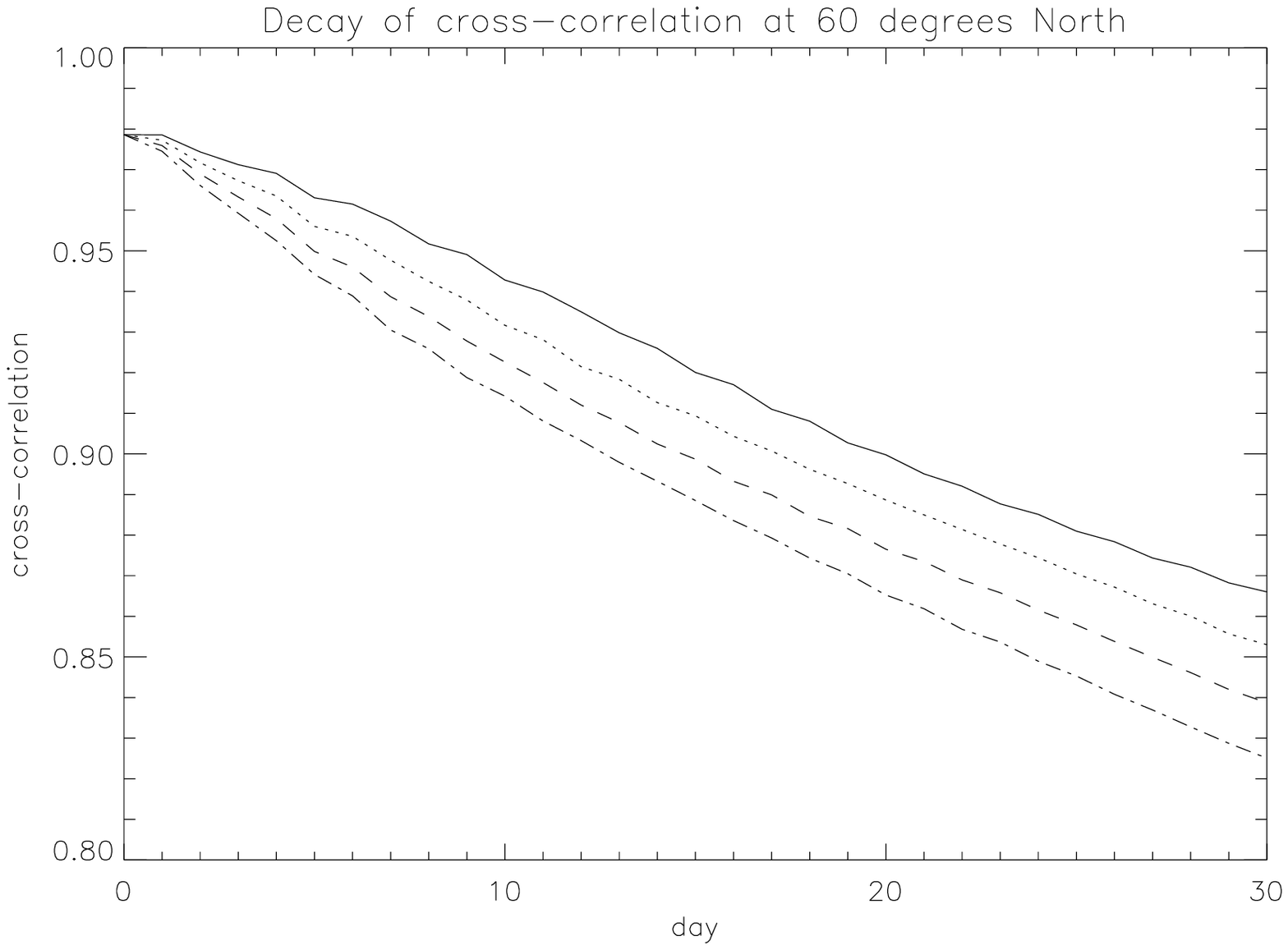,angle=0,width=7.0cm} \\
\end{tabular}
\caption{Cross-correlation of calculated radial field with the
observed radial field for 1995, latitude 30\degr (left) and 60\degr
(right).  The solid line represents $\eta=250$km$^{2}$s$^{-1}$, the
dotted line $\eta=350$km$^{2}$s$^{-1}$, the dashed line
$\eta=450$km$^{2}$s$^{-1}$ and the dot-dash line
$\eta=550$km$^{2}$s$^{-1}$.  We should not expect a cross-correlation
of exactly 1 at day 0 since we are correlating an observed field with
a calculated one.}
\label{cc95}
\end{figure*}

From three of Maxwell's equations 
$$
\nabla{\times}{\bf B}=\mu{\bf j},
$$
$$
\nabla.{\bf B}=0,
$$
$$
\nabla{\times}{\bf E}=-\frac{\partial{\bf B}}{\partial t}
$$ 
and Ohm's Law 
$$
{\bf j}=\sigma({\bf E}+{\bf v}{\times}{\bf B}),
$$ 
where $\sigma$ is the conductivity we get the induction
equation
\begin{equation}
\frac{\partial {\bf B}}{\partial
t}=\nabla{\times}({\bf v}{\times}{\bf B})-\nabla{\times}{\bf E'}.
\end{equation}
Here ${\bf E'}$ is given by
$$
E'_{r}=\frac{\eta}{r\sin\theta}\left[\frac{\partial}{\partial\theta}
(B_{\phi}\sin\theta)-\frac{\partial B_{\theta}}{\partial\phi}\right],
$$
$$
E'_{\theta}=\frac{\eta}{r\sin\theta}\frac{\partial B_{r}}{\partial\phi},
$$
$$
E'_{\phi}=-\frac{\eta}{r}\frac{\partial B_{r}}{\partial\theta},
$$
with $\eta$=1/$\mu\sigma$ being the magnetic diffusivity
\cite{vanballegooijen98}.  This assumes that there is no radial
transport of the magnetic field, and that the meridional flow is
poleward and the same as the solar value given by
\begin{equation}
u(\lambda) = \left\{ \begin{array}{ll}
-u_{0}\sin(\frac{\pi\lambda}{\lambda_{0}}) & \mbox{if
$|\lambda|<\lambda_{0}$} 
\\
0 & \mbox{otherwise}.
\end{array}
\right. 
\end{equation}
Here $\lambda\equiv\frac{\pi}{2}-\theta$ is the latitude,
$\lambda_{0}$ gives the latitude above which the meridional flow is
zero, $\lambda_{0}=75\degr$ and $u_{0}=11$m s$^{-1}$ which is close to
the predicted value \cite{kitch99}.  The values of $B_{lm}$, $A_{lm}$
and $J_{lm}$ are evolved using the induction equation according to the
meridional flow, the observed differential rotation and using various
values of the magnetic diffusion, ranging from 250 to 550
km$^{2}$s$^{-1}$ ({\em cf.} the solar value of 450 km$^{2}$s$^{-1}$).

The differential rotation is of the form 
\begin{equation}
\Omega(\theta)=12.2434-0.0564\cos^{2}\theta  \mbox{ rad d }^{-1}
\end{equation}
where $\Omega$ is the rotation rate \cite{donati97doppler}.  We have
assumed that $\eta$ is uniform across the surface, although there is
evidence that this may not be the case for the Sun \cite{berger98}.

\subsection{Calculating the evolved field}

\begin{figure*}
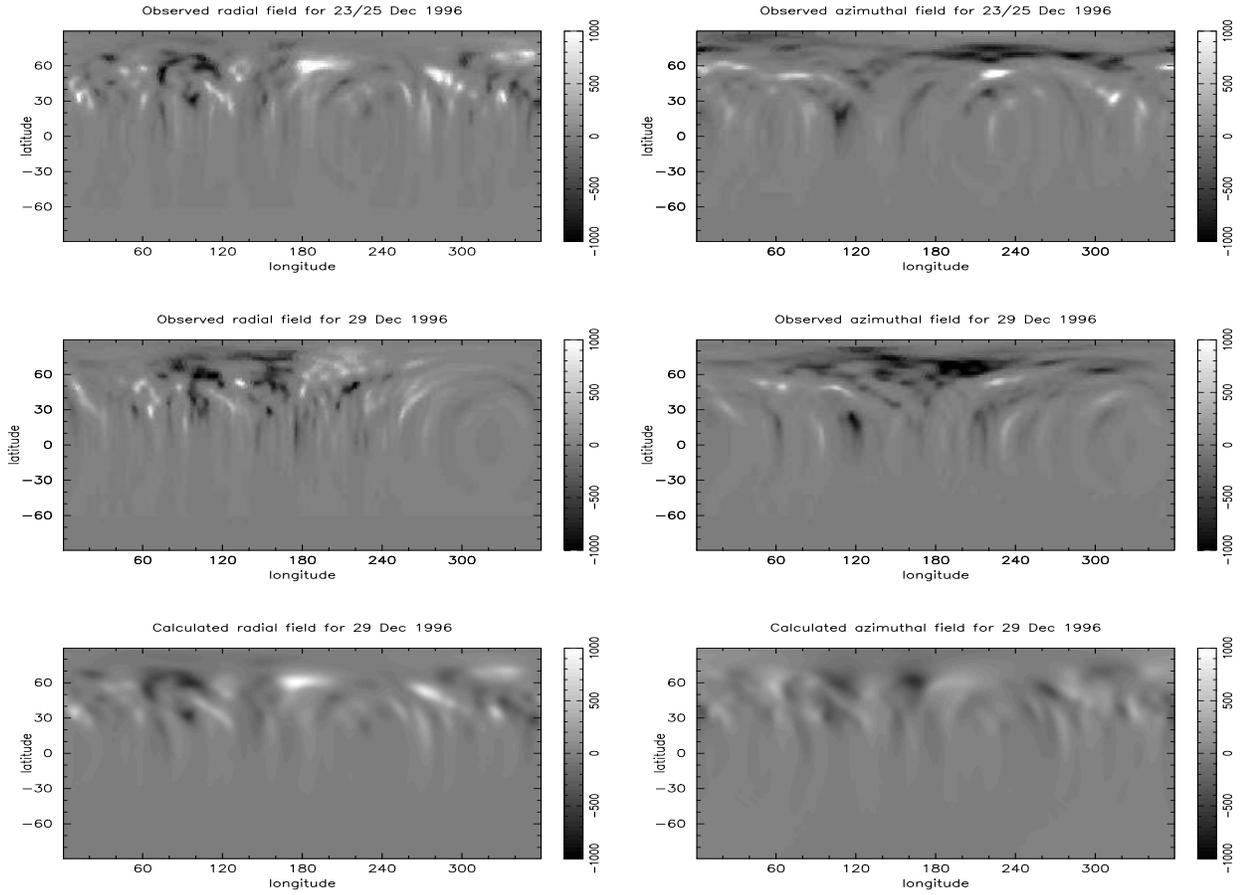

\begin{tabular}{cc}
\psfig{file=fig03a.eps,angle=270,width=8.0cm,height=4.0cm}
&
\psfig{file=fig03b.eps,angle=270,width=8.0cm,height=4.0cm}
\\
\psfig{file=fig03c.eps,angle=270,width=8.0cm,height=4.0cm}
&
\psfig{file=fig03d.eps,angle=270,width=8.0cm,height=4.0cm}
\\
\psfig{file=fig03e.eps,angle=270,width=8.0cm,height=4.0cm}
&
\psfig{file=fig03f.eps,angle=270,width=8.0cm,height=4.0cm}
\\
\end{tabular}
\caption{This diagram shows the azimuthal and radial components of the
field in 1996. The top row gives the observed radial (left) and azimuthal 
(right) fields for Dec 23/25. The middle row gives the observed radial
(left) and azimuthal (right) fields for Dec 29. The bottom row gives the
calculated radial (left) and azimuthal (right) fields for Dec 29, assuming
that $\eta$=450km$^{2}$s$^{-1}$}
\label{field96}
\end{figure*}

The third stage is to take the evolved coefficients $B_{lm}$,$A_{lm}$
and $J_{lm}$ and the associated Legendre functions $P_{lm}$ and
calculate the three components of the magnetic field- $B_{r}$
(radial), $B_{\phi}$ (azimuthal) and $B_{\theta}$ (meridional)- from
them.  These will be given by
$$
B_{r}(r,\theta,\phi)=\sum_{l=1}^{N}\sum_{m=-l}^{l}B_{lm}P_{lm}(\theta)e^{im\phi}
$$
$$
B_{\theta}(r,\theta,\phi)=\sum_{l=1}^{N}\sum_{m=-l}^{l}\lambda_{l}\left[-A_{lm}(r)
\frac{dP_{lm}}{d\theta}+iJ_{lm}\frac{mP_{lm}}{\sin\theta}\right]e^{im\phi}
$$
$$
B_{\phi}(r,\theta,\phi)=\sum_{l=1}^{N}\sum_{m=-l}^{l}\lambda_{l}\left[-A_{lm}(r)
\frac{imP_{lm}}{\sin\theta}-J_{lm}\frac{dP_{lm}}{d\theta}\right]e^{im\phi}
$$
where 
$$
\lambda_{l}=\frac{1}{l(l+1)}.
$$

\section{Results}

As an initial consistency check we computed the evolution of the
magnetic energy in the field at the surface, calculating the ratio of
magnetic energy in the evolved case to the original.

A magnetic field has energy $B^{2}/2\mu$ per unit volume, so the total
energy is $$W=\int_{volume}\frac{B^{2}}{2\mu}dV$$ Fig \ref{magen}
shows this for the 1998 field.  The left-hand panel shows the degree
of diffusive decay of the field energy over 30 days.  The right-hand
panel shows the increase in energy that would occur in the absence of
diffusion, due to the winding-up of the field by the differential
rotation.

\subsection{Long-term evolution of the surface field}

We took the observed field for the 1995 and 1998 observations and
evolved it over 30 days according to the induction equation (4).  This
allowed us to study the variation with time of the cross-correlation
of the radial component observed on the first night with that
calculated for subsequent nights (Fig \ref{cc95}).  We chose two
latitudes: 30 and 60\degr north.  The results for 1998 are
qualitatively similar.  In all cases, the cross-correlation function
decays by approximately 10\% over 30 days.  Although choosing a higher
value for the diffusivity does cause a more rapid decay of the field
and hence a faster decay of the cross-correlation function, it is
still not enough to explain the complete lack of correlation found by
\scite{barnes98alpper} for He699.  It appears that for AB Dorad\^{u}s,
if diffusion and differential rotation were the only processes causing
the field to evolve, that even after one month there should still be a
good correlation.

\begin{figure*} 
\begin{tabular}{cc}
\psfig{file=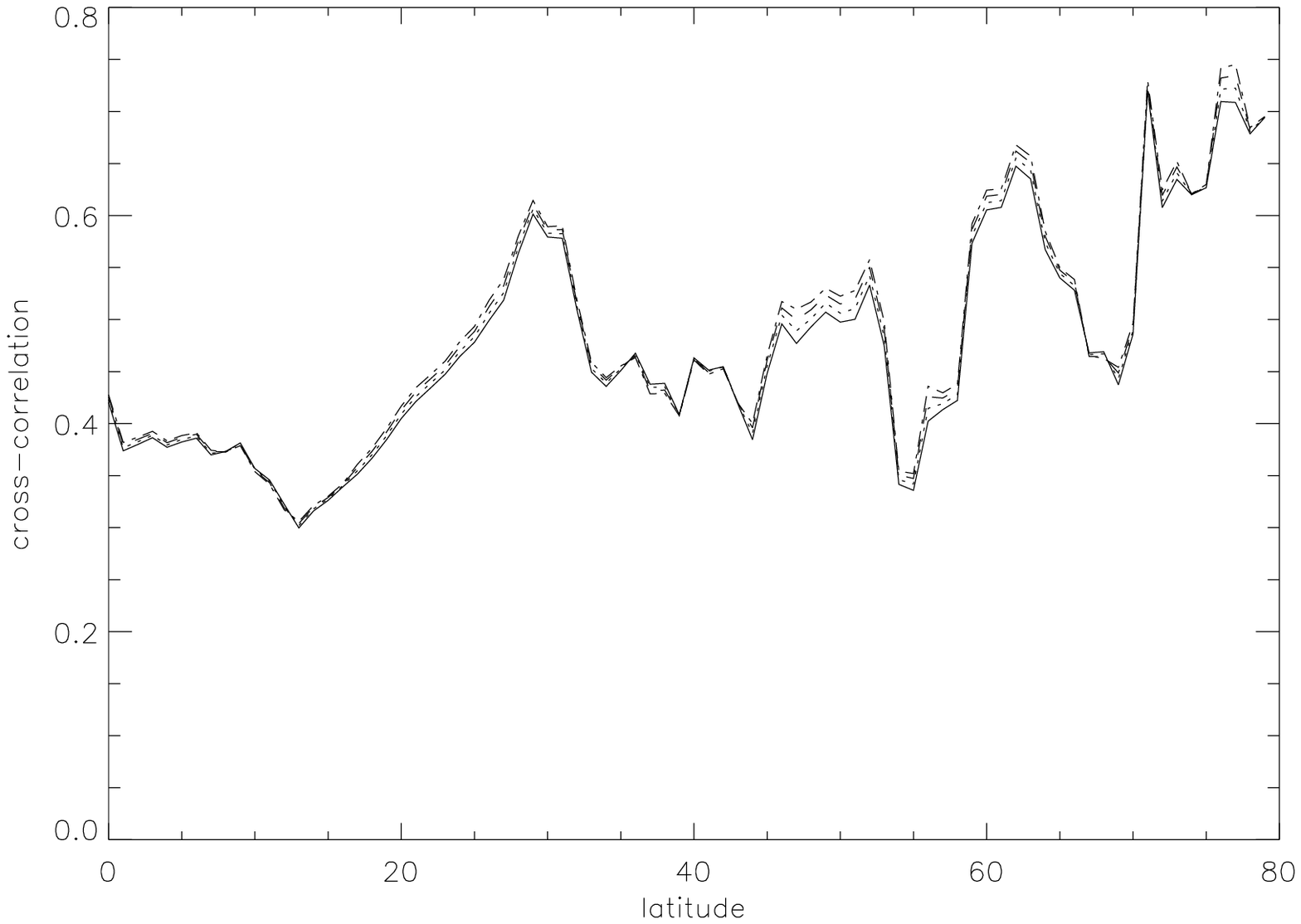,angle=0,width=7.0cm}
&
\psfig{file=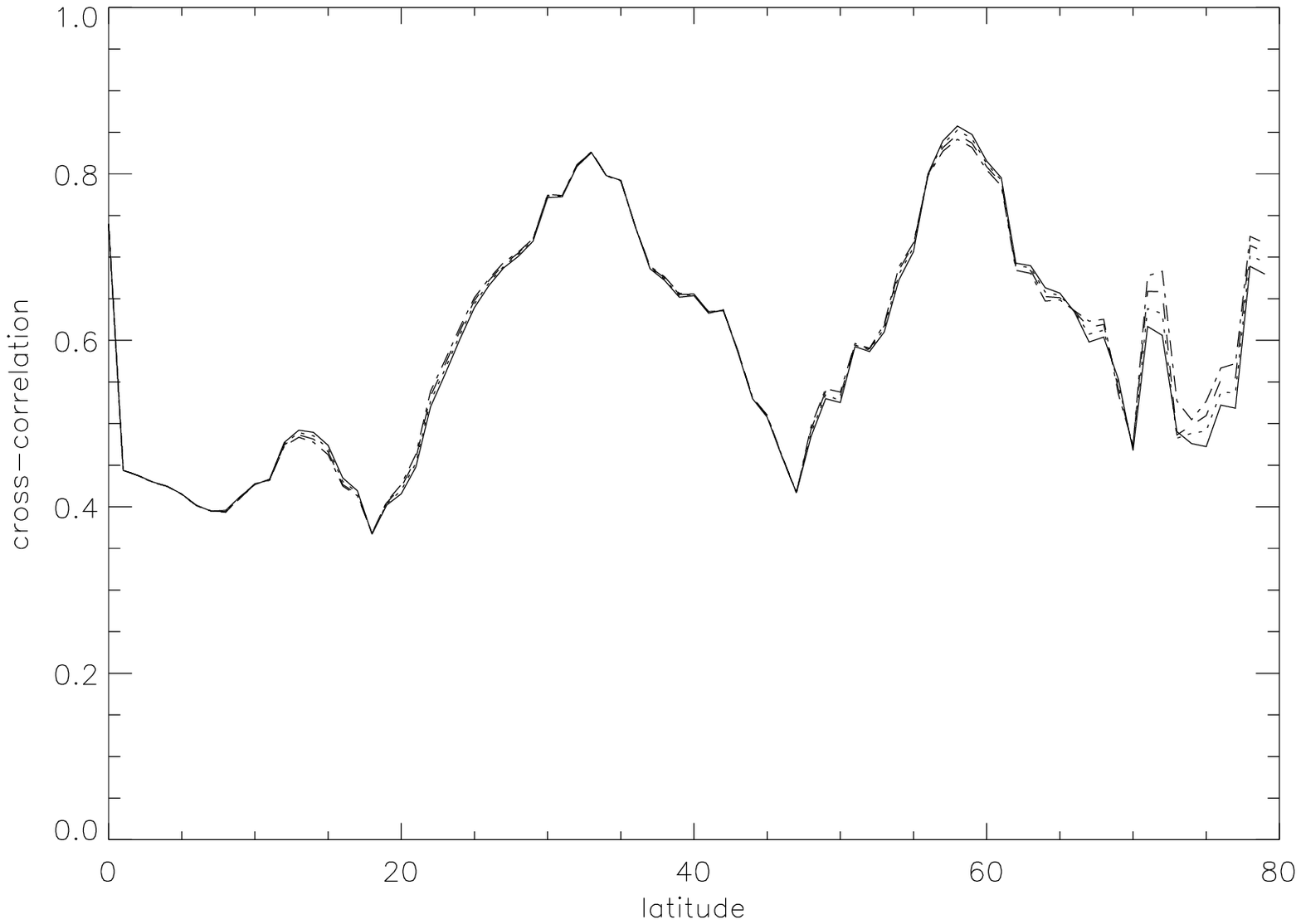,angle=0,width=7.0cm}
\\
\end{tabular}
\caption{The peak amplitude of the cross-correlation function between
the observed and calculated radial (left) and azimuthal (right) fields
for latitudes between 0 and 80\degr north for 1996 Dec 29.  The solid
line represents $\eta=250$km$^{2}$s$^{-1}$, the dotted line
$\eta=350$km$^{2}$s$^{-1}$, the dashed line $\eta=450$km$^{2}$s$^{-1}$
and the dot-dash line $\eta=550$km$^{2}$s$^{-1}$.}
\label{compare}
\end{figure*}

\begin{figure*}
\begin{tabular}{cc}
\psfig{file=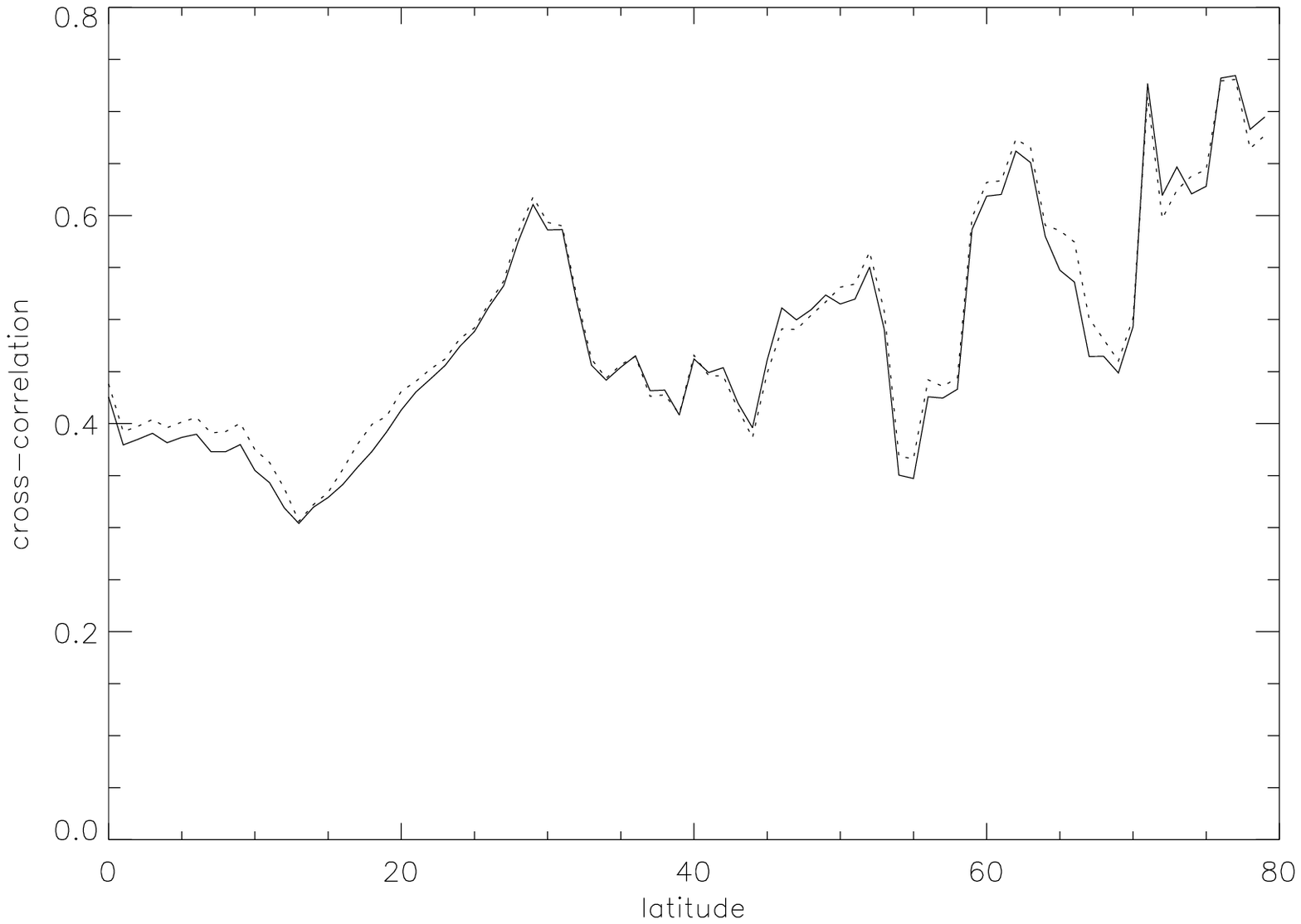,angle=0,width=7.0cm}
&
\psfig{file=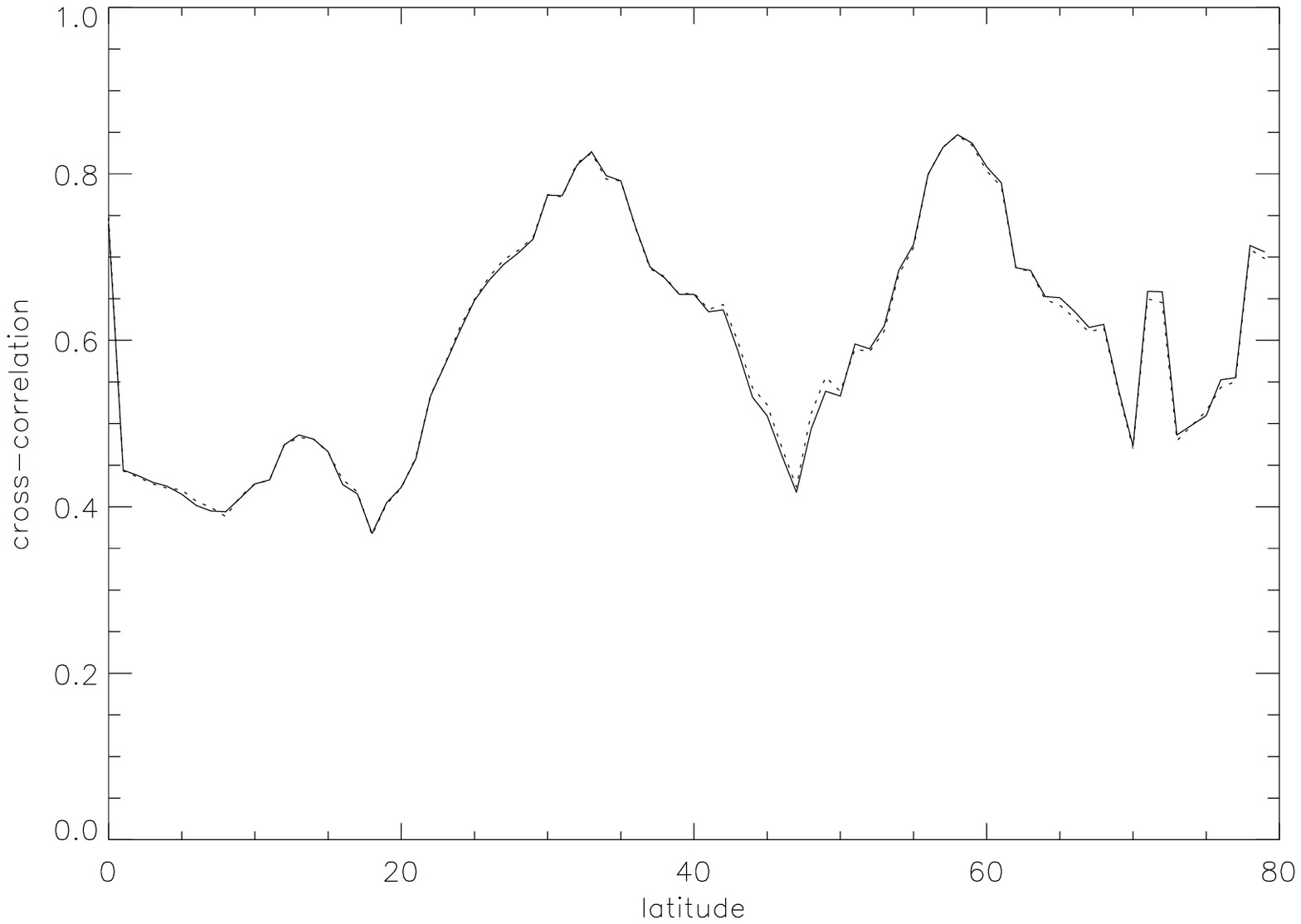,angle=0,width=7.0cm}
\\
\end{tabular}
\caption{The peak amplitude of the cross-correlation function between
the observed and calculated radial (left) and azimuthal (right) fields
for latitudes between 0 and 80\degr north for 1996 Dec 29 with
$\eta=450$km$^{2}$s$^{-1}$.  The solid line is with the differential
rotation, and the dotted line is without the differential rotation.}
\label{nocompare}
\end{figure*}

\subsection{Short-term evolution}

For comparison, we can look at the field evolution over a much shorter
period of time.  This has the advantage that we can compare our
results with the observed evolution of the field during a single
observing run.  Here we use data from the 1996 run, in which we
secured two sets of magnetic maps, separated by 5 nights
(Fig.~\ref{field96}).  Our aim is to determine whether the observed
magnetic elements retain their identities over a period of 5 nights in
the presence of diffusion and differential rotation.

We began by investigating the effect of varying the diffusivity.  We
evolved the 1996 Dec 23/25 field forward in time using various values
of the diffusion coefficient.  In each case we cross-correlated the
resulting radial field map with the observed radial field for Dec 29
(Fig \ref{compare}).  We ensured that the cross-correlation only
involved those longitudes that were well-observed, {\em viz.}
18-180\degr \cite{jardine99potent}.  The cross-correlation function
was computed for each of a set of latitudes between 0 and 80\degr in
the visible hemisphere, and the amplitude of the strongest peak 
in the ccf at each latitude is plotted in Fig.~\ref{compare}. It can 
be seen that the five-day span of these observations is too short for 
differences in the value chosen for diffusivity to have much effect.

We also considered the case where the field was allowed to evolve 
under the influence of diffusion only, i.e. with the differential 
rotation switched off. We found that over the 5-day span of the 1996 
December observations, the influence of the differential rotation was 
negligible (Fig.~\ref{nocompare}).

From these we see that altering the value of the diffusion and
removing the differential rotation have little effect on the
cross-correlation over 5 nights.  We would need observations over a
longer timescale, say a month, to be able to look for meaningful
results.  
 
\section{The high-latitude azimuthal field}

\scite{jardine99potent} demonstrated that the high-latitude azimuthal
band of field was not reproduced by modelling the field as potential. 
Here we investigate whether the differential rotation could produce
this band by taking the 1998 radial field map and adding in a cap of
unipolar radial field extending from the pole to latitude 80\degr,
well within the dark polar spot seen in the stellar surface-brightness
map.  An identical cap of opposite polarity was added to the
unobservable hemisphere to conserve flux.  We emphasize that the
polarity and strength of the dark polar cap cannot be determined
directly from the observations, since the low surface brightness and
strong foreshortening suppress the Zeeman signal from this part of the
star.  For each of a range of plausible polar field strengths and
polarities we evolved the field forward in time for 5 days, then
compared the mean value of $B_{\phi}$ at each latitude with the
observed value.

\begin{figure}
\centering
\psfig{file=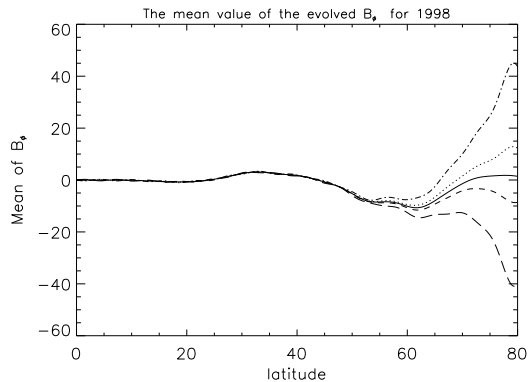,angle=0,width=7.0cm}
\caption{Azimuthally-averaged values of $B_{\phi}$ as a function of
latitude computed from the 1998 data, evolved for 5 nights with
$\eta=250$km$^{2}$s$^{-1}$ and with different values of the polar field. 
The solid line is the case where the polar field is set to 0G, the
dotted line where it is set to +1000G, the dashed line where it is set
to -1000G, the dash-dot line where it is set to +4000G and the
long-dashed line where it is set to -4000G.}
\label{negcap98}
\end{figure}

\begin{figure}
\centering
\psfig{file=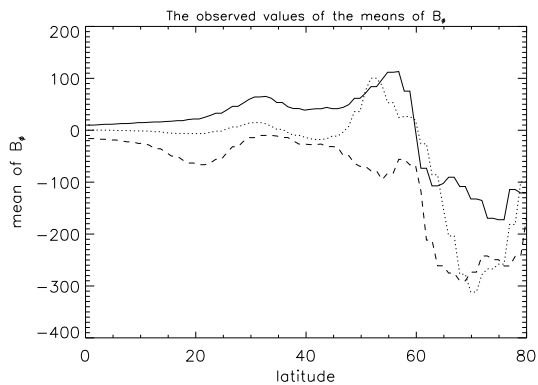,angle=0,width=7.0cm} 
\caption{Azimuthally-averaged values of $B_{\phi}$ as a function of
latitude in the observed field maps
for 1995 (solid line), 1996 Dec 23/25 (dotted line) and 1998 (dashed line).}
\label{torcompare}
\end{figure}

We see from Fig.~\ref{negcap98} that a high-latitude azimuthal band of
negative polarity is produced when we shear the image at the observed
differential rotation rate with a polar cap having $B_{r}<0$ and vice
versa.  The results confirm that the observed differential rotation is
capable of producing a high-latitude negative azimuthal band, as is
seen in the observations from all three observing seasons
(Fig.~\ref{torcompare}).  This predicts that the magnetic polarity of
the dark polar region was predominantly negative in all three seasons.

The latitude of the maximum in $B_{\phi}$ occurs at the edge of the
imposed unipolar cap in Fig.~\ref{negcap98}.  The azimuthal field is
localised here because most field lines originating in the north polar
cap are connected to high northern latitudes just outside the cap. 
The direction of these field lines depends strongly upon the
distribution of radial field at mid to high latitudes.  Changing the
strength of the polar cap has little effect on field lines at low
latitude.  This corresponds to our result that $\sum B_{\phi}$ varies
little at low latitude as the strength of the cap is altered (as seen
in Fig \ref{negcap98}).

The observed azimuthal band plotted in Fig.~\ref{torcompare} is more
diffuse than the model, having a broader peak between latitudes
65\degr and 80\degr in all three seasons' data.  This corresponds
roughly to the edge of the dark polar region, which extends to
latitude 70\degr or so.  The breadth of the peak suggests a more
gradual fall-off in the polar field than we imposed on the model.  The
apparent decrease in field strength at latitudes above 70\degr can
probably be ascribed to suppression of the Zeeman signal within the
dark polar region.

The strength of the azimuthal field after only 5 nights is
substantially less than that observed.  This is not surprising, since
the timescale on which shear can generate an azimuthal field with a
strength comparable to the radial and meridional field will be of
order the equator-pole lap time of 110 days.  The diffusion time for
length scales comparable to the size of the individual magnetic
regions in the images is also of this order.  Once an equilibrium is
established between diffusion and differential rotation, we would
expect the azimuthal field strength to be an order of magnitude
greater than that produced after 5 days, in agreement with the
observations.  On a 100-day timescale, however, we expect the picture
to be complicated further by the emergence of new flux, making a
direct comparison with the observations problematic.

\section{Conclusions}

We have modelled the evolution of the magnetic field of AB Dorad\^{u}s
due to the effects of differential rotation and diffusion.  We use as
a starting point the Zeeman Doppler images obtained on three
consecutive years and assume that the field is initially potential,
but evolves away from this state as a function of time.  

Over a timescale of 20 to 30 days we have determined, as a function of
time, the cross-correlation of our model radial magnetic field with
the observed radial component on the first night.  We find that over
one month, the cross-correlation function decays by about 10\%. 
Observations of He699 by \scite{barnes98alpper} show however that
cross-correlating the observed spot distributions over this timescale
gives much more rapid decrease of the cross-correlation function. 
This result suggests that the evolution of AB Dorad\^{u}s's surface
magnetic field is not governed solely by diffusion and differential
rotation.  We conclude that these results are more likely to be due to
the effects of flux emergence changing the spot distribution than the
effects of diffusion or differential rotation.

This result is independent of the assumed degree of field diffusion. We
have compared the effects of values of $\eta$ ranging from 250 to 450
km$^{2}$s$^{-1}$ and found the results to be qualitatively the same. The
presence of some diffusion is of course necessary (and we have confirmed
that the magnetic energy grows monotonically with time in the absence of
diffusion). The exact value of $\eta$ seems however to have little effect 
on the five-day timescale of a typical observing run.
Since diffusion has little effect on the flux distribution, the
differential rotation acts simply to advect the field. Consequently,
although at each latitude the peak of the cross-correlation function may
be at a different longitude \cite{donati97doppler}, its actual
value is virtually unchanged by the effects of differential rotation. 

We have also compared the radial and azimuthal magnetic fields
generated by our model over 5 nights with those obtained from Zeeman
Doppler images on 1996 Dec 23/25 and 29, and found the agreement to be
excellent.  The evolution of the azimuthal field is of particular
interest with regard to the band of high latitude unidirectional
azimuthal field seen in the Zeeman-Doppler images.  We have
investigated whether the shearing effect of the differential rotation
is sufficient to generate this band of field.  The polarity of this
band depends upon the sign of the radial field in the polar cap,
whereas its strength depends on the competition between shear and
diffusion.  Since the diffusion timescale for resolvable features is
comparable to the winding time, we expect the azimuthal field to
attain a strength comparable to the radial and meridional field near
the boundary of the polar cap, as is indeed observed.  Our results
suggest that the differential rotation could play a major part in the
creation and preservation of a high latitude azimuthal band.
  
\section{Acknowledgements}
We would like to thank Drs.  D. Mackay, A. van Ballegooijen and M.
Ferreira for useful discussions and assistance during the course of
this work.  We also thank Drs.  G. Hussain and L. Kitchatinov for
their careful reading of the manuscript and offering comments.  GRP
acknowledges the support of a studentship from the University of St
Andrews.

%\bibliographystyle{mn}
%\bibliography{iau_journals,master,ownrefs,mythesis,gaiteecs12}

\end{document}